\documentclass[twocolumn,amsmath,amssymb,aps,preprintnumbers,showpacs,superscriptaddress]{revtex4}

\usepackage{bm}         
\usepackage{amssymb}
\usepackage{amsmath}
\usepackage{graphics}
\usepackage{epsfig}
\usepackage{color}
\usepackage{ulem}

\begin{document}

\title{The phase boundary of superconducting niobium thin films with antidot arrays fabricated with microsphere photolithography}

\author{D.~Bothner}
\address{Physikalisches Institut and Center for Collective Quantum Phenomena in LISA$^+$, Universit\"{a}t T\"{u}bingen, Auf der Morgenstelle 14, D-72076 T\"{u}bingen, Germany}
\author{C.~Clauss}
\address{1. Physikalisches Institut, Universit\"{a}t Stuttgart, Pfaffenwaldring 57, D-70550 Stuttgart, Germany}
\author{E.~Koroknay}
\affiliation{Institut f\"{u}r Halbleiteroptik und Funktionelle Grenzfl\"{a}chen and Research Center SCoPE, Universit\"{a}t Stuttgart, Allmandring 3, D-70569 Stuttgart, Germany}
\author{M.~Kemmler}
\affiliation{Physikalisches Institut and Center for Collective Quantum Phenomena in LISA$^+$, Universit\"{a}t T\"{u}bingen, Auf der Morgenstelle 14, D-72076 T\"{u}bingen, Germany}
\author{T.~Gaber}
\affiliation{Physikalisches Institut and Center for Collective Quantum Phenomena in LISA$^+$, Universit\"{a}t T\"{u}bingen, Auf der Morgenstelle 14, D-72076 T\"{u}bingen, Germany}
\author{M.~Jetter}
\affiliation{Institut f\"{u}r Halbleiteroptik und Funktionelle Grenzfl\"{a}chen and Research Center SCoPE, Universit\"{a}t Stuttgart, Allmandring 3, D-70569 Stuttgart, Germany}
\author{M.~Scheffler}
\affiliation{1. Physikalisches Institut, Universit\"{a}t Stuttgart, Pfaffenwaldring 57, D-70550 Stuttgart, Germany}
\author{P.~Michler}
\affiliation{Institut f\"{u}r Halbleiteroptik und Funktionelle Grenzfl\"{a}chen and Research Center SCoPE, Universit\"{a}t Stuttgart, Allmandring 3, D-70569 Stuttgart, Germany}
\author{M.~Dressel}
\affiliation{1. Physikalisches Institut, Universit\"{a}t Stuttgart, Pfaffenwaldring 57, D-70550 Stuttgart, Germany}
\author{D.~Koelle}
\affiliation{Physikalisches Institut and Center for Collective Quantum Phenomena in LISA$^+$, Universit\"{a}t T\"{u}bingen, Auf der Morgenstelle 14, D-72076 T\"{u}bingen, Germany}
\author{R.~Kleiner}
\affiliation{Physikalisches Institut and Center for Collective Quantum Phenomena in LISA$^+$, Universit\"{a}t T\"{u}bingen, Auf der Morgenstelle 14, D-72076 T\"{u}bingen, Germany}
\date{\today}

\begin{abstract}

The experimental investigation of the $I_c(B)$--$T_c(B)$ phase boundary of superconducting niobium films with large area quasihexagonal hole arrays is reported. 
The hole arrays were patterned with microsphere photolithography.
We investigate the perforated niobium films by means of electrical directed current transport measurements close to the transition temperature $T_c$ in perpendicularly applied magnetic fields.
We find pronounced modulations of the critcal current with applied magnetic field, which we interpret as a consequence of commensurable states between the Abrikosov vortex lattice and the quasihexagonal pinning array.
Furthermore, we observe Little-Parks oscillations in the critical temperature vs magnetic field.

\end{abstract}


\maketitle

\begin{figure}
\centering {\includegraphics{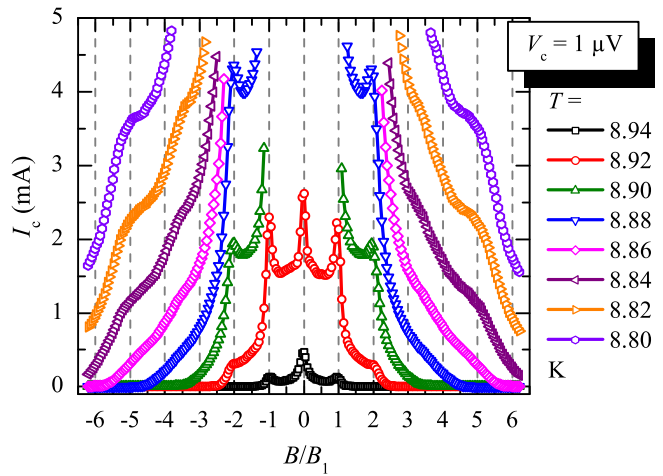}}
\caption{(Color online) Critical current $I_c$ vs magnetic flux density $B$ of a superconducting thin film with a quasihexagonal lattice of antidots; different curves correspond to different temperatures $T$. The flux density axis is normalized to the first matching flux density $B_1\approx 3.4\,$mT.}
\label{fig:Graph3}
\end{figure}

\begin{figure}
\centering {\includegraphics{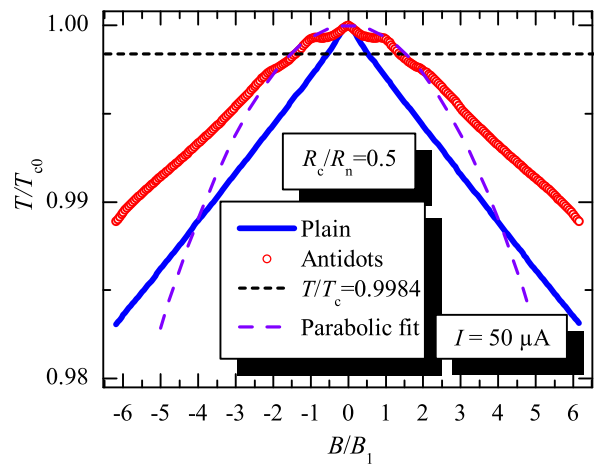}}
\caption{(Color online) Reduced critical temperature $T_c/T_{\textrm{c0}}$ vs normalized magnetic flux density $B/B_1$ of a plain superconducting thin film (blue line) and a sample with a quasihexagonal lattice of antidots (red circles). The horizontal dashed line represents the temperature, at which $\xi(T_W/T_{\textrm{c0}})=W$, the dashed parabola is a fit to the three data points of the antidot sample at $B=0$, $B=B_1$ and $B=-B_1$.}
\label{fig:Graph4}
\end{figure}

Nowadays superconducting thin films are used for a huge variety of superconducting microelectronic devices such as Josephson junctions, superconducting quantum interference devices (SQUIDs) and coplanar waveguide resonators.
Typically, these thin films are made of type-II superconductors and are penetrated by quantized magnetic flux when operated in magnetic fields or when biased with sufficiently high currents.
The investigation of these magnetic flux lines (Abrikosov vortices) and their individual and collective interactions with natural and artificial defects in the superconductor is of high interest and subject to many experimental and theoretical studies for several decades now.
One reason for this sustained scientific attention is that unpinned Abrikosov vortices respond with a dissipative motion to any current flowing in their vicinity.
In many cases this motion is directly related to a reduction of the performance (increased noise, lowered quality factor, shortened coherence time) of the microelectronic device.
Defects, however, act as local energy minima and pinning sites and are able to reduce or even completely suppress vortex motion and the related dissipation \cite{Fiory78, Martin97, Moshchalkov98, Haeffner09, Song09}.
For instance, it has been demonstrated that the flux noise in SQUIDs and the dissipation in coplanar microwave resonators can be reduced by strategically positioned microholes (antidots) \cite{Selders00, Bothner11}.
A second and more fundamental point is associated with the fact, that an ensemble of Abrikosov vortices interacting with an ensemble of defects in the superconductor constitutes a highly designable and controllable model system for repulsively interacting particles in a twodimensional potential landscape.
In such systems it is possible to investigate static effects such as the formation of quasicrystals \cite{Misko05, Kemmler06, Villegas06, Kramer09, Misko10} or the controlled introduction of potential landscape disorder \cite{Reichhardt07, Kemmler09, Rosen10} as well as dynamic effects such as mode locking phenomena \cite{Martinoli75, VanLook99, Kokubo02} and ratchet dynamics \cite{Villegas03, Silva06}.
Of particular interest in both research branches is the case, when the typical length scales of the defect topology, i.e. size and mutual distance, are comparable to the intrinsic length scales of the superconductor, that are the coherence length $\xi$ and the magnetic penetration depth $\lambda$, which are both temperature dependent and diverge at the critical temperature $T_c$.
Well below the critical temperature, $\lambda$ and $\xi$ can usually be found in the micro- to nanometer range.
To pattern large areas of superconducting films with submicron-scaled high density arrays of defects constitutes a non-trivial challenge to standard optical (limited by resolution) or electron beam (limited by time) lithography.
\begin{figure}[h]
\centering {\includegraphics{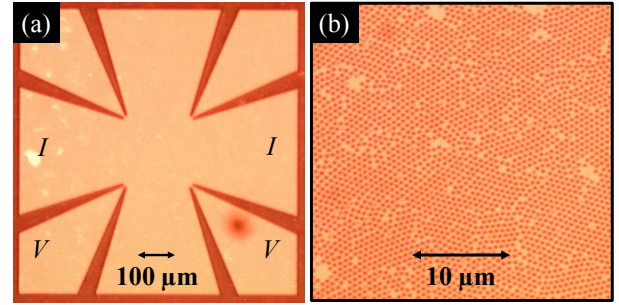}}
\caption{(Color online) (a) Layout of a $800\times800$\,$\mu$m$^2$ large cross shaped bridge structure with a square center area of $200\times200$\,$\mu$m$^2$ for the four-probe current voltage characterization of superconducting thin films with pinning landscapes; (b) zoom into (a), niobium film with a microsphere patterned quasihexagonal array of antidots.}
\label{fig:Graph1}
\end{figure}
\begin{figure*}
\centering {\includegraphics{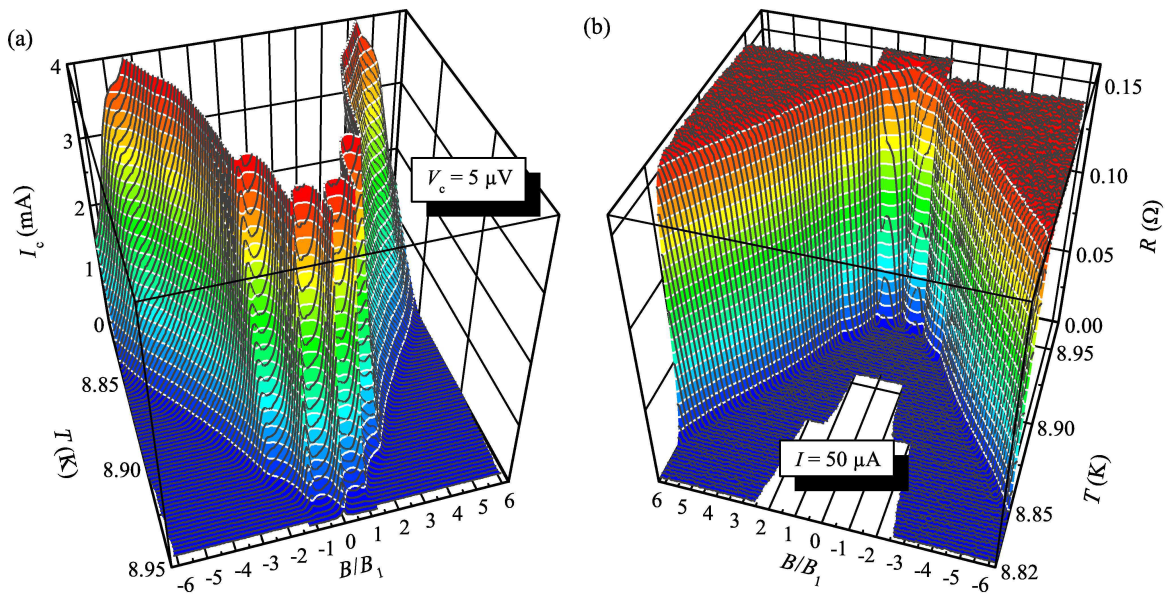}}
\caption{(Color online) (a) Critical current $I_c$ vs temperature $T$ and magnetic flux density $B$ of a superconducting thin film with a quasihexagonal lattice of antidots for a voltage criterion of $V_c = 5\,\mu$V; (b) Resistance $R$ of the niobium film vs $T$ and $B$ measured with a current $I=50\,\mu$A. Flux density axes are normalized to the first matching flux density $B_1\approx 3.4\,$mT.}
\label{fig:Graph2}
\end{figure*}
It has been demonstrated with different approaches, that taking advantage of self-assembling structures can provide a way out of the difficulties to cover large areas with tiny structures on reasonable timescales.
The techniques involved vary from using anodized aluminum as substrate material \cite{Welp02} over depositing the superconducting film on a layer of microspheres \cite{Vinckx06} to generating structures by inverse diblock copolymer micelle formation \cite{Eisenmenger07}.
These fabrication techniques are limited to certain substrate materials or they induce changes in the substrate properties and/or the properties of the superconductor.
Here we adopt another method to fabricate large area quasihexagonal arrays of submicron sized antidots, which is independent of the substrate material and does not influence the superconducting material more than any standard lithography process \cite{Wu08}.
In a previous study we have demonstrated, that with this fabrication technique it is possible to reduce the vortex associated losses in superconducting microwave resonators by more than one order of magnitude \cite{Bothner12}.
In the present manuscript we analyze the properties of our microsphere patterned Nb thin films by means of transport measurements close to the transition temperature with a particular focus on signatures of commensurabilities between the antidot and vortex lattices.
We also investigate a transition between the wire network (width of the superconducting material between the antidots $W<\xi(T)$) and the thin film regime with vortices ($W>\xi(T)$) in our samples.
We fabricated our samples by first depositing a $t = 150\,$nm thick niobium film on a r-cut sapphire wafer by dc magnetron sputtering.
Afterwards we cut the wafer into individual chips and carried out the lithography steps.
For the fabrication of perforated samples the chips were covered with photoresist and on top of that with a monolayer of water suspended polystyrene colloids in a Langmuir-Blodgett deposition process.
The microspheres have a diameter of $D_s = 770 (\pm25)\,$nm and act as a self-assembled array of UV-light focusing microlenses, leading to a quasihexagonal hole array after the exposure, their removal and the resist development.
For a perfect, hexagonal close-packed array with $D_s = 770\,$nm one would get a corresponding hole density of $n_h \approx 1.95\,\mu \textrm{m}^{-2}$.
In reality, however, one could expect deviations from the ideal packaging due to disorder during the self-assembling.
After transferring the hole array into the Nb film via reactive ion etching (SF$_6$) we patterned cross-shaped bridge structures for electric transport characterizations into the films.
For this we used standard optical shadow-mask lithography and another SF$_6$ reactive ion etching step.
Figure~\ref{fig:Graph1} (a) shows one of the bridge structures with a square center area of $200\times200\,\mu$m${^2}$.
The antidots have an approximate diameter $D_a=370\,$nm, which in principle can be easily varied by adjusting the lithography exposure time.
In Fig.~\ref{fig:Graph1} (b) a zoom-in to the niobium film with antidots is depicted, which shows a domain-like pattern of holes with some blemishes.
Due to the cross shape of our bridges, the transport current in the center area is not homogeneous but somewhat spreadened.
For ratchet devices, it has been discussed that such a sample geometry may strongly affect experimental results \cite{Gonzalez07, Silhanek08, Gonzalez08}.
Also in our case, this geometry might have an effect on the absolute values of the measured quantities, such as the critical current $I_c$, what e.g. impedes a precise determination of critical current densities.
However, the results presented in this manuscript do neither sensitively depend on the absolute values of the measured quantities and on the local direction of the transport current.
We have also patterned and characterized bridges with $100\times100\,\mu$m${^2}$ and $50\times50\,\mu$m${^2}$ large center squares, but the experimental results showed no dependence on the bridge size.
Note, that the niobium chips were taken from the same wafer as the chips for our previous study on resonators \cite{Bothner12}.
To characterize our samples we mount them into a low-temperature setup, that provides a temperature stability $\Delta T < 1\,$mK, and contact them electrically with wire bonds.
We apply a magnetic field perpendicular to the film plane using a superconducting coil, and monitor current voltage characteristics (IVCs) for many values of magnetic field and temperature.
After collecting all IVCs we extract the desired information as the critical current $I_c$, the critical temperature $T_c$ or the resistance $R$ vs magnetic flux density $B$ and temperature $T$.
We choose the threshold voltage $V_c$ defining $I_c$ and the measurement current $I$ for the resistance $R$ during the evaluation.
To reduce the voltage noise, we take several IVCs at each value for $B$ and $T$ and post-process the raw data (averaging and smoothing), such that we are more sensitive to modulations of the $I_c$--$T_c$ phase boundary with respect to the applied magnetic flux.
As an overview of the whole phase boundary Fig.~\ref{fig:Graph2} depicts (a) the critical current $I_c$ and (b) the resistance $R$ vs magnetic flux density and temperature of one of our samples close to $T_c$.
Obviously, the phase boundary, i.e. the critical current $I_c(B)$, the critical temperature $T_c(B)$ and the resistance $R(B)$, boundary is non-monotonous.
All of the quantities strongly modulate with the applied flux density.
There are several ``canyons'' and ``ridges'' indicating commensurate states between the flux line lattice and the hole array.
For a more detailed view of the position of the maxima and minima in the 3D phase boundary, it is convenient to extract single data slices.
Figure~\ref{fig:Graph3} shows several individual curves for the critical current $I_c(B)$ corresponding to vertical cuts for constant temperatures through Fig.~\ref{fig:Graph2} (a).
The flux density axis in Fig.~\ref{fig:Graph2} and Fig.~\ref{fig:Graph3} is normalized to the first pronounced maximum in the critical current $B_1\approx3.4\,$mT.
Under the assumption, that $B_1$ corresponds to equal densities of vortices and holes in the sample, we find an antidot density of $n_a=B/\phi_0\approx1.65\,\mu$m$^{-2}$, which is somewhat smaller than the previously calculated $n_h\approx1.95\,\mu$m$^{-2}$ for a lattice without any defects.
This difference is most likely due to defects and dislocations in the pinning array, cf. also Fig.~\ref{fig:Graph1} (b).
We of course can not be sure, that $B_1$ indeed corresponds to equal vortex and antidot densities, as previous theoretical and experimental studies on randomly diluted and disordered triangular antidot arrays have shown a certain variability of the matching fields \cite{Reichhardt07, Kemmler09, Rosen10}.
The same studies however suggest, that the amplitudes of the maxima get smaller and the peaks smear out, when the dilution/disorder in the pinning array is increased.
The observation of quite strong and sharp maxima in the phase boundary of our samples at $B=B_1$ and also at $B=B_2=2B_1$ thus suggest, that we have a rather ordered lattice in accordance with the impression from optical images (cf. Fig~\ref{fig:Graph1}).
We believe, that the data of Fig.~\ref{fig:Graph3} also show fingerprints of the disorder in the lattice for higher field values $B>B_2$.
For a perfect triangular lattice, one would expect maxima of the critical current predominantly at integer multiples of $B_1$.
However, for higher values we find two shoulders around $B\approx3.5B_1$ and $B\approx5B_1$ and none at $B=3B_1$ or $B=4B_1$.
This observation might reflect the necessity of additional vortices at interstitial positions to stabilize the vortex lattice (cf. also \cite{Reichhardt07, Kemmler09}), leading to robust configurations at $B\approx3.5B_1$ and $B\approx5B_1$.
The effect of shifted and missing peaks is related to the widely discussed phenomenon of pinning interstitial vortices by ''caging`` them between regularly pinned ones.
The presence of interstitial vortices can induce an increased maximum number of vortices, which can occupy the pinning sites.
This in turn can cause an increase of the collective pinning strength by adding vortices.
These effects have been observed in theoretical and experimental studies on periodic as well as quasiperiodic pinning arrays \cite{Silhanek05, Berdiyorov06, Misko10, Rablen11, Cao11, Latimer12}.
In all of these cases, the caged vortices have a spatial distribution with the same symmetry as the underlying pinning lattice.
In the present study, however, we assume the interstitial vortices to be filling up the imperfect antidot lattice at the positions of missing antidots, what is very similar to the situation in randomly diluted pinning arrays \cite{Reichhardt07, Kemmler09}.
Thus, the flux values, at which stable vortex configurations and matching features in the phase boundary appear, are more related to the completeness and quality of the pinning array than to the intrinsic symmetry of the domain-like array parts.
Finally, although not fully identical, all these mechanisms are closely related to each other.
Besides analyzing the phase boundary by taking horizontal slices, we can also take vertical slices for chosen currents or voltages of the 3D boundary and end up with the critical temperature vs magnetic field plots (or second critical magnetic field vs temperature, respectively) for different $T_c$ ($B_{\textrm{c2}}$) criteria.
Figure~\ref{fig:Graph4} shows a plot of $T_c(B)/T_c(0)=T_c/T_{\textrm{c0}}$ of a perforated sample (symbols) for a resistance criterion $R_c/R_n=0.5$ and measured with an applied current  $I=50\,\mu$A.
$R_n$ denotes the normal state resistance at $T=10\,$K.
Oscillations of the critical temperature with the applied flux are clearly visible, which we associate with Little-Parks oscillations \cite{Little62} as already observed in many studies on superconducting wire networks and thin films with pinning arrays before \cite{Pannetier84, Behrooz86, Nori87, Patel07}.
We also plot the critical temperature vs magnetic field of a plain reference sample in Fig.~\ref{fig:Graph4} (a) and calculate from these data the coherence length $\xi(0)=16\,$nm of the niobium by fitting it to the bulk expression $B_{\textrm{c2}}=\phi_0/[2\pi\xi(T)^2]$ with $\xi(T)=\xi(0)(1-T/T_c)^{-1/2}$ \cite{Tinkham}.
As the coherence length is significantly smaller than the BCS coherence length for niobium $\xi_0=39\,$nm our films are in the ``dirty limit'' with a free mean path of $l=1.37\xi(0)^2/\xi_0=9\,$nm.
Assuming that the coherence length is not changed by the antidot patterning process and remembering, that the minimal width of superconducting material between two holes is $W=D_s-D_a=400\,$nm, we calculate the reduced temperature, at which $\xi(T_W)=W$, to $T_W/T_{\textrm{c0}}=0.9984$.
Above $T_W$, which is marked with a dashed horizontal line in Fig.~\ref{fig:Graph4}, the superconductivity in our system can be viewed as one-dimensional and the thin film expression $B_{\textrm{c2}}=\sqrt{12}\phi_0/[2\pi W\xi(T)]$ should apply \cite{Tinkham, Welp02}.
This gives a parabolic dependence $T_c/T_{\textrm{c0}}\propto B^2$ for $T/T_{\textrm{c0}}>0.9984$, what seems to be in reasonable agreement with our data.
$T_c/T_{\textrm{c0}}$ of the perforated sample indeed appears like a Little-Parks modulated nonlinear background close to $T_{\textrm{c0}}$ with a development to a nearly linear behaviour for smaller temperatures $T/T_{\textrm{c0}}<0.9984$.
By fitting the three single points at $B=0$, $B=B_1$ and $B=-B_1$ to the above parabolic expression and using $W=400\,$nm, we calculate a coherence length $\xi(0)\approx18.2\,$nm in good agreement with the number extracted from the plain sample, although it is probably somewhat overestimated.
The effective remaining superconductor width between the holes $W$ is certainly larger than the used minimal value of $400\,$nm for two reasons. 
First, the holes have a circular shape and second, some of them are missing.
So it might even be, that the coherence length in the perforated sample is somewhat smaller than in the plain, what would be supported by the impression of a slightly smaller slope in the more linear regime.
A fit of the data for the perforated sample in this region however is difficult, as there are still modulations due to the antidots superimposed.
In summary we have investigated the $I_c(B)-T_c(B)$ phase boundary of superconducting Nb films, which were patterned with quasihexagonal arrays of submicron sized antidots.
This fabrication method is especially interesting for microwave devices as it does not change the substrate and superconductor properties significantly.
Our experiments revealed signatures of both, order and disorder in the pinning lattice, which was patterned by using a monolayer of self-assembling polystyrene colloids as microlenses for optical lithography.
Pronounced sharp maxima of the critical current in the first and second matching field indicate a high ordering of the pinning sites, whereas the shoulder-like structures at noninteger higher field values might be related to pinning lattice blemishes and disorder.
We also observe Little-Parks oscillations of the critical temperature and approximately identify the wire network to thin film transition in our samples.
The coherence length of the perforated film, extracted from the critical temperature vs magnetic field dependence, is in good agreement with that of a plain film, confirming that our patterning method has hardly influenced the properties of the niobium.
We have performed experiments close to $T_c$, but when using superconductors with a higher magnetic penetration depth, the results are also relevant at temperatures $T=4.2\,$K or even in the mK regime.
This situation is given in very thin or dirty superconducting films and in different superconducting materials such as YBCO or NbN.
In principle our patterning technique can be used with even smaller spheres, which would lead to commensurability effects at much lower temperatures and higher magnetic fields.
In these cases, the critical current and the pinning efficiency will modulate with the applied field similar to the presented manner, what has to be considered for the design of possible devices.
This work has been supported by the Deutsche Forschungsgemeinschaft via the SFB/TRR 21 and by the European Research Council via SOCATHES.
DB gratefully acknowledges support by the Evangelisches Studienwerk e.V. Villigst.
MK gratefully acknowledges support by the Carl-Zeiss Stiftung.


\begin{thebibliography}{26}
\expandafter\ifx\csname natexlab\endcsname\relax\def\natexlab#1{#1}\fi
\expandafter\ifx\csname bibnamefont\endcsname\relax
  \def\bibnamefont#1{#1}\fi
\expandafter\ifx\csname bibfnamefont\endcsname\relax
  \def\bibfnamefont#1{#1}\fi
\expandafter\ifx\csname citenamefont\endcsname\relax
  \def\citenamefont#1{#1}\fi
\expandafter\ifx\csname url\endcsname\relax
  \def\url#1{\texttt{#1}}\fi
\expandafter\ifx\csname urlprefix\endcsname\relax\def\urlprefix{URL }\fi
\providecommand{\bibinfo}[2]{#2}
\providecommand{\eprint}[2][]{\url{#2}}

\bibitem[{\citenamefont{Fiory et~al.}(1978)\citenamefont{Fiory, Hebard, and
  Minnich}}]{Fiory78}
\bibinfo{author}{\bibfnamefont{A.~T.} \bibnamefont{Fiory}},
  \bibinfo{author}{\bibfnamefont{A.~F.} \bibnamefont{Hebard}},
  \bibnamefont{and} \bibinfo{author}{\bibfnamefont{R.~P.}
  \bibnamefont{Minnich}}, \bibinfo{journal}{J. Phys. Colloques}
  \textbf{\bibinfo{volume}{39}}, \bibinfo{pages}{633} (\bibinfo{year}{1978}).
   
\bibitem[{\citenamefont{Mart\'in et~al.}(1997)\citenamefont{Mart\'in, V\'elez, Nogu\'es, and Schuller}}]{Martin97}
\bibinfo{author}{\bibfnamefont{J.~I.}~\bibnamefont{Mart\'in}},
  \bibinfo{author}{\bibfnamefont{M.}~\bibnamefont{Vel\'ez}},
  \bibinfo{author}{\bibfnamefont{J.}~\bibnamefont{Nogu\'es}}, \bibnamefont{and}
  \bibinfo{author}{\bibfnamefont{I.~K.}~\bibnamefont{Schuller}},
  \bibinfo{journal}{Phys. Rev. Lett.} \textbf{\bibinfo{volume}{79}},
  \bibinfo{pages}{1929} (\bibinfo{year}{1997}).
   
\bibitem[{\citenamefont{Moshchalkov et~al.}(1998)\citenamefont{Moshchalkov,
  Baert, Metlushko, Rosseel, Bael, Temst, Bruynseraede, and
  Jonckheere}}]{Moshchalkov98}
\bibinfo{author}{\bibfnamefont{V.~V.} \bibnamefont{Moshchalkov}},
  \bibinfo{author}{\bibfnamefont{M.}~\bibnamefont{Baert}},
  \bibinfo{author}{\bibfnamefont{V.~V.} \bibnamefont{Metlushko}},
  \bibinfo{author}{\bibfnamefont{E.}~\bibnamefont{Rosseel}},
  \bibinfo{author}{\bibfnamefont{M.~J.~V.}~\bibnamefont{Bael}},
  \bibinfo{author}{\bibfnamefont{K.}~\bibnamefont{Temst}},
  \bibinfo{author}{\bibfnamefont{Y.}~\bibnamefont{Bruynseraede}},
  \bibnamefont{and}
  \bibinfo{author}{\bibfnamefont{R.}~\bibnamefont{Jonckheere}},
  \bibinfo{journal}{Phys. Rev. B} \textbf{\bibinfo{volume}{57}},
  \bibinfo{pages}{3615} (\bibinfo{year}{1998}).
  
\bibitem[{\citenamefont{H\"affner et~al.}(2009)\citenamefont{H\"affner, Kemmler, L\"offler, Vega G\'omez, Fleischer, Kleiner, Koelle, and Kern}}]{Haeffner09}
\bibinfo{author}{\bibfnamefont{M.}~\bibnamefont{H\"affner}},
  \bibinfo{author}{\bibfnamefont{M.}~\bibnamefont{Kemmler}},
  \bibinfo{author}{\bibfnamefont{R.}~\bibnamefont{L\"offler}},
  \bibinfo{author}{\bibfnamefont{B.}~\bibnamefont{Vega G\'omez}},
  \bibinfo{author}{\bibfnamefont{M.}~\bibnamefont{Fleischer}},
  \bibinfo{author}{\bibfnamefont{R.}~\bibnamefont{Kleiner}},
  \bibinfo{author}{\bibfnamefont{D.}~\bibnamefont{Koelle}}, \bibnamefont{and}
  \bibinfo{author}{\bibfnamefont{D.~P.}~\bibnamefont{Kern}},
  \bibinfo{journal}{Microelectronic Engineering} \textbf{\bibinfo{volume}{86}},
  \bibinfo{pages}{147003} (\bibinfo{year}{2009}).
  
\bibitem[{\citenamefont{Song et~al.}(2009{\natexlab{b}})\citenamefont{Song,
  DeFeo, Yu, and Plourde}}]{Song09}
\bibinfo{author}{\bibfnamefont{C.}~\bibnamefont{Song}},
  \bibinfo{author}{\bibfnamefont{M.~P.} \bibnamefont{DeFeo}},
  \bibinfo{author}{\bibfnamefont{K.}~\bibnamefont{Yu}}, \bibnamefont{and}
  \bibinfo{author}{\bibfnamefont{B.~L.~T.} \bibnamefont{Plourde}},
  \bibinfo{journal}{Appl. Phys. Lett.} \textbf{\bibinfo{volume}{95}},
  \bibinfo{pages}{232501} (\bibinfo{year}{2009}{\natexlab{b}}).

\bibitem[{\citenamefont{Selders and W\"ordenweber}(2000)\citenamefont{Selders and W\"ordenweber}}]{Selders00}
\bibinfo{author}{\bibfnamefont{P.}~\bibnamefont{Selders}}
  \bibnamefont{and} \bibinfo{author}{\bibfnamefont{R.} \bibnamefont{W\"ordenweber}},
  \bibinfo{journal}{Appl. Phys. Lett.} \textbf{\bibinfo{volume}{76}},
  \bibinfo{pages}{3277} (\bibinfo{year}{2000}).
  
\bibitem[{\citenamefont{Bothner et~al.}(2011)\citenamefont{Bothner,
  Gaber, Kemmler, Koelle, and Kleiner}}]{Bothner11}
\bibinfo{author}{\bibfnamefont{D.}~\bibnamefont{Bothner}},
  \bibinfo{author}{\bibfnamefont{T.}~\bibnamefont{Gaber}},
  \bibinfo{author}{\bibfnamefont{M.}~\bibnamefont{Kemmler}},
  \bibinfo{author}{\bibfnamefont{D.}~\bibnamefont{Koelle}},
  \bibnamefont{and} \bibinfo{author}{\bibfnamefont{R.} \bibnamefont{Kleiner}},
  \bibinfo{journal}{Appl. Phys. Lett.} \textbf{\bibinfo{volume}{98}},
  \bibinfo{pages}{102504} (\bibinfo{year}{2011}).
  
\bibitem[{\citenamefont{Misko et~al.}(2005)\citenamefont{Misko, Savel'ev and Nori}}]{Misko05}
\bibinfo{author}{\bibfnamefont{V.}~\bibnamefont{Misko}},
  \bibinfo{author}{\bibfnamefont{S.}~\bibnamefont{Savel'ev}}, \bibnamefont{and}
  \bibinfo{author}{\bibfnamefont{F.}~\bibnamefont{Nori}},
  \bibinfo{journal}{Phys. Rev. Lett.} \textbf{\bibinfo{volume}{95}},
  \bibinfo{pages}{177007} (\bibinfo{year}{2005}).
  
\bibitem[{\citenamefont{Kemmler et~al.}(2006)\citenamefont{Kemmler, G\"urlich, Sterck, P\"ohler, Neuhaus, Siegel, Kleiner and Koelle}}]{Kemmler06}
\bibinfo{author}{\bibfnamefont{M.}~\bibnamefont{Kemmler}},
  \bibinfo{author}{\bibfnamefont{C.}~\bibnamefont{G\"urlich}},
  \bibinfo{author}{\bibfnamefont{A.}~\bibnamefont{Sterck}},
  \bibinfo{author}{\bibfnamefont{H.}~\bibnamefont{P\"ohler}},
  \bibinfo{author}{\bibfnamefont{M.}~\bibnamefont{Neuhaus}},
  \bibinfo{author}{\bibfnamefont{M.}~\bibnamefont{Siegel}},
  \bibinfo{author}{\bibfnamefont{R.}~\bibnamefont{Kleiner}}, \bibnamefont{and}
  \bibinfo{author}{\bibfnamefont{D.}~\bibnamefont{Koelle}},
  \bibinfo{journal}{Phys. Rev. Lett.} \textbf{\bibinfo{volume}{97}},
  \bibinfo{pages}{147003} (\bibinfo{year}{2006}).
  
\bibitem[{\citenamefont{Villegas et~al.}(2006)\citenamefont{Villegas, Montero, Li, and Schuller}}]{Villegas06}
\bibinfo{author}{\bibfnamefont{J.~E.}~\bibnamefont{Villegas}},
  \bibinfo{author}{\bibfnamefont{M.~I.}~\bibnamefont{Montero}},
  \bibinfo{author}{\bibfnamefont{C.-P.}~\bibnamefont{Li}}, \bibnamefont{and}
  \bibinfo{author}{\bibfnamefont{I.~K.}~\bibnamefont{Schuller}},
  \bibinfo{journal}{Phys. Rev. Lett.} \textbf{\bibinfo{volume}{97}},
  \bibinfo{pages}{027002} (\bibinfo{year}{2006}).
  
\bibitem[{\citenamefont{Kramer et~al.}(2009)\citenamefont{Kramer, Silhanek, Van de Vondel, Raes and Moshchalkov}}]{Kramer09}
\bibinfo{author}{\bibfnamefont{R.~B.~G.}~\bibnamefont{Kramer}},
  \bibinfo{author}{\bibfnamefont{A.~V.}~\bibnamefont{Silhanek}},
  \bibinfo{author}{\bibfnamefont{J.}~\bibnamefont{Van de Vondel}},
  \bibinfo{author}{\bibfnamefont{B.}~\bibnamefont{Raes}}, \bibnamefont{and}
  \bibinfo{author}{\bibfnamefont{V.~V.}~\bibnamefont{Moshchalkov}},
  \bibinfo{journal}{Phys. Rev. Lett.} \textbf{\bibinfo{volume}{103}},
  \bibinfo{pages}{067007} (\bibinfo{year}{2009}).
  
\bibitem[{\citenamefont{Misko et al.}(2010)\citenamefont{Misko, Bothner, Kemmler, Kleiner, Koelle, Peeters, and Nori}}]{Misko10}
\bibinfo{author}{\bibfnamefont{V.~R.} \bibnamefont{Misko}},
  \bibinfo{author}{\bibfnamefont{D.}~\bibnamefont{Bothner}},
  \bibinfo{author}{\bibfnamefont{M.}~\bibnamefont{Kemmler}},
  \bibinfo{author}{\bibfnamefont{R.}~\bibnamefont{Kleiner}},
  \bibinfo{author}{\bibfnamefont{D.}~\bibnamefont{Koelle}},
  \bibinfo{author}{\bibfnamefont{F.~M.}~\bibnamefont{Peeters}}, \bibnamefont{and}
  \bibinfo{author}{\bibfnamefont{F.} \bibnamefont{Nori}},
  \bibinfo{journal}{Phys. Rev. B} \textbf{\bibinfo{volume}{82}},
  \bibinfo{pages}{184512} (\bibinfo{year}{2010}).
  
\bibitem[{\citenamefont{Reichhardt and Olson Reichhardt}(2007)}]{Reichhardt07}
  \bibinfo{author}{\bibfnamefont{C.}~\bibnamefont{Reichhardt}},
  \bibnamefont{and} \bibinfo{author}{\bibfnamefont{C.~J.} \bibnamefont{Olson Reichhardt}},
  \bibinfo{journal}{Phys. Rev. B} \textbf{\bibinfo{volume}{76}},
  \bibinfo{pages}{094512} (\bibinfo{year}{2007}).
  
\bibitem[{\citenamefont{Kemmler et~al.}(2009)\citenamefont{Kemmler, Bothner,
  Ilin, Siegel, Kleiner, and Koelle}}]{Kemmler09}
  \bibinfo{author}{\bibfnamefont{M.}~\bibnamefont{Kemmler}},
  \bibinfo{author}{\bibfnamefont{D.}~\bibnamefont{Bothner}},
  \bibinfo{author}{\bibfnamefont{K.}~\bibnamefont{Ilin}},
  \bibinfo{author}{\bibfnamefont{M.}~\bibnamefont{Siegel}},
  \bibinfo{author}{\bibfnamefont{R.}~\bibnamefont{Kleiner}},
  \bibnamefont{and} \bibinfo{author}{\bibfnamefont{D.} \bibnamefont{Koelle}},
  \bibinfo{journal}{Phys. Rev. B} \textbf{\bibinfo{volume}{79}},
  \bibinfo{pages}{184509} (\bibinfo{year}{2009}).
  
\bibitem[{\citenamefont{Rosen et~al.}(2010)\citenamefont{Rosen, Sharoni, and Schuller}}]{Rosen10}
  \bibinfo{author}{\bibfnamefont{Y.~J.}~\bibnamefont{Rosen}},
  \bibinfo{author}{\bibfnamefont{A.}~\bibnamefont{Sharoni}},
  \bibnamefont{and} \bibinfo{author}{\bibfnamefont{I.~K.} \bibnamefont{Schuller}},
  \bibinfo{journal}{Phys. Rev. B} \textbf{\bibinfo{volume}{82}},
  \bibinfo{pages}{014509} (\bibinfo{year}{2010}).

\bibitem[{\citenamefont{Martinoli et~al.}(1975)\citenamefont{Martinoli, Daldini, Leemann and Stocker}}]{Martinoli75}
  \bibinfo{author}{\bibfnamefont{P.}~\bibnamefont{Martinoli}},
  \bibinfo{author}{\bibfnamefont{O.}~\bibnamefont{Daldini}},
  \bibinfo{author}{\bibfnamefont{C.}~\bibnamefont{Leemann}},
  \bibnamefont{and} \bibinfo{author}{\bibfnamefont{E.} \bibnamefont{Stocker}},
  \bibinfo{journal}{Solid State Commun.} \textbf{\bibinfo{volume}{17}},
  \bibinfo{pages}{205} (\bibinfo{year}{1975}).
  
\bibitem[{\citenamefont{Van Look et~al.}(1999)\citenamefont{Van Look, Rosseel, Van Bael, Temst, Moshchalkov and Bruynseraede}}]{VanLook99}
  \bibinfo{author}{\bibfnamefont{L.}~\bibnamefont{Van Look}},
  \bibinfo{author}{\bibfnamefont{E.}~\bibnamefont{Rosseel}},
  \bibinfo{author}{\bibfnamefont{M.~J.}~\bibnamefont{Van Bael}},
  \bibinfo{author}{\bibfnamefont{K.}~\bibnamefont{Temst}},
  \bibinfo{author}{\bibfnamefont{V.~V.}~\bibnamefont{Moshchalkov}},
  \bibnamefont{and} \bibinfo{author}{\bibfnamefont{Y.} \bibnamefont{Bruynseraede}},
  \bibinfo{journal}{Phys. Rev. B} \textbf{\bibinfo{volume}{60}},
  \bibinfo{pages}{R6998} (\bibinfo{year}{1999}).
  
\bibitem[{\citenamefont{Kokubo et~al.}(2002)\citenamefont{Kokubo, Besseling, Vinokur and Kes}}]{Kokubo02}
  \bibinfo{author}{\bibfnamefont{N.}~\bibnamefont{Kokubo}},
  \bibinfo{author}{\bibfnamefont{R.}~\bibnamefont{Besseling}},
  \bibinfo{author}{\bibfnamefont{V.~M.}~\bibnamefont{Vinokur}},
  \bibnamefont{and} \bibinfo{author}{\bibfnamefont{P.~H.}~\bibnamefont{Kes}},
  \bibinfo{journal}{Phys. Rev. Lett.} \textbf{\bibinfo{volume}{88}},
  \bibinfo{pages}{247004} (\bibinfo{year}{2002}).  
  
\bibitem[{\citenamefont{Villegas et~al.}(2003)\citenamefont{Villegas, Savel'ev, Nori, Gonzalez, Anguita, Garc\'ia and Vicent}}]{Villegas03}
  \bibinfo{author}{\bibfnamefont{J.~E.}~\bibnamefont{Villegas}},
  \bibinfo{author}{\bibfnamefont{S.}~\bibnamefont{Savel'ev}},
  \bibinfo{author}{\bibfnamefont{F.}~\bibnamefont{Nori}},
  \bibinfo{author}{\bibfnamefont{E.~M.}~\bibnamefont{Gonzalez}},
  \bibinfo{author}{\bibfnamefont{J.~V.}~\bibnamefont{Anguita}},
  \bibinfo{author}{\bibfnamefont{R.}~\bibnamefont{Garc\'ia}},
  \bibnamefont{and} \bibinfo{author}{\bibfnamefont{J.~L.} \bibnamefont{Vicent}},
  \bibinfo{journal}{Science} \textbf{\bibinfo{volume}{302}},
  \bibinfo{pages}{1188-1191} (\bibinfo{year}{2003}). 

\bibitem[{\citenamefont{de Souza Silva et~al.}(2006)\citenamefont{de Souza Silva, Van de Vondel, Morelle and Moshchalkov}}]{Silva06}
  \bibinfo{author}{\bibfnamefont{C.~C.}~\bibnamefont{de Souza Silva}},
  \bibinfo{author}{\bibfnamefont{J.}~\bibnamefont{Van de Vondel}},
  \bibinfo{author}{\bibfnamefont{M.}~\bibnamefont{Morelle}},
  \bibnamefont{and} \bibinfo{author}{\bibfnamefont{V.~V.} \bibnamefont{Moshchalkov}},
  \bibinfo{journal}{Nature} \textbf{\bibinfo{volume}{440}},
  \bibinfo{pages}{651-654} (\bibinfo{year}{2006}).

\bibitem[{\citenamefont{Welp et~al.}(2002)\citenamefont{Welp,
  Xiao, Jiang, Vlasko-Vlasov, Bader, Crabtree, Liang, Chik, and Xu}}]{Welp02}
\bibinfo{author}{\bibfnamefont{U.}~\bibnamefont{Welp}},
  \bibinfo{author}{\bibfnamefont{Z.~L.}~\bibnamefont{Xiao}},
  \bibinfo{author}{\bibfnamefont{J.~S.}~\bibnamefont{Jiang}},
  \bibinfo{author}{\bibfnamefont{V.~K.}~\bibnamefont{Vlasko-Vlasov}},
  \bibinfo{author}{\bibfnamefont{S.~D.}~\bibnamefont{Bader}},
  \bibinfo{author}{\bibfnamefont{G.~W.}~\bibnamefont{Crabtree}},
  \bibinfo{author}{\bibfnamefont{J.}~\bibnamefont{Liang}},
  \bibinfo{author}{\bibfnamefont{H.}~\bibnamefont{Chik}},
  \bibnamefont{and} \bibinfo{author}{\bibfnamefont{J.~M.} \bibnamefont{Xu}},
  \bibinfo{journal}{Phys. Rev. B} \textbf{\bibinfo{volume}{66}},
  \bibinfo{pages}{212507} (\bibinfo{year}{2002}).

\bibitem[{\citenamefont{Vinckx et~al.}(2006)\citenamefont{Vinckx,
  Vanacken, and Moshchalkov}}]{Vinckx06}
\bibinfo{author}{\bibfnamefont{W.}~\bibnamefont{Vinckx}},
  \bibinfo{author}{\bibfnamefont{J.}~\bibnamefont{Vanacken}},
  \bibnamefont{and} \bibinfo{author}{\bibfnamefont{V.~V.} \bibnamefont{Moshchalkov}},
  \bibinfo{journal}{J. Appl. Phys.} \textbf{\bibinfo{volume}{100}},
  \bibinfo{pages}{044307} (\bibinfo{year}{2006}).

\bibitem[{\citenamefont{Eisenmenger et~al.}(2007)\citenamefont{Eisenmenger,
  Oettinger, Pfahler, Plettl, Walther, and Ziemann}}]{Eisenmenger07}
\bibinfo{author}{\bibfnamefont{J.}~\bibnamefont{Eisenmenger}},
  \bibinfo{author}{\bibfnamefont{M.}~\bibnamefont{Oettinger}},
  \bibinfo{author}{\bibfnamefont{C.}~\bibnamefont{Pfahler}},
  \bibinfo{author}{\bibfnamefont{A.}~\bibnamefont{Plettl}},
  \bibinfo{author}{\bibfnamefont{P.}~\bibnamefont{Walther}},
  \bibnamefont{and} \bibinfo{author}{\bibfnamefont{P.} \bibnamefont{Ziemann}},
  \bibinfo{journal}{Phys. Rev. B} \textbf{\bibinfo{volume}{75}},
  \bibinfo{pages}{144514} (\bibinfo{year}{2007}).
   
\bibitem[{\citenamefont{Wu et~al.}(2008)\citenamefont{Wu,
  Dey, Memis, Katsnelson, and Mohseni}}]{Wu08}
\bibinfo{author}{\bibfnamefont{W.}~\bibnamefont{Wu}},
  \bibinfo{author}{\bibfnamefont{D.}~\bibnamefont{Dey}},
  \bibinfo{author}{\bibfnamefont{O.~G.}~\bibnamefont{Memis}},
  \bibinfo{author}{\bibfnamefont{A.}~\bibnamefont{Katsnelson}},
  \bibnamefont{and} \bibinfo{author}{\bibfnamefont{H.} \bibnamefont{Mohseni}},
  \bibinfo{journal}{Nanoscale Res. Lett.} \textbf{\bibinfo{volume}{3}},
  \bibinfo{pages}{351-354} (\bibinfo{year}{2008}).
  
\bibitem[{\citenamefont{Bothner et~al.}(2012)\citenamefont{Bothner,
  Clauss, Koroknay, Kemmler, Gaber, Jetter, Scheffler, Michler, Dressel, Koelle, and Kleiner}}]{Bothner12}
\bibinfo{author}{\bibfnamefont{D.}~\bibnamefont{Bothner}},
  \bibinfo{author}{\bibfnamefont{C.}~\bibnamefont{Clauss}},
  \bibinfo{author}{\bibfnamefont{E.}~\bibnamefont{Koroknay}},
  \bibinfo{author}{\bibfnamefont{M.}~\bibnamefont{Kemmler}},
  \bibinfo{author}{\bibfnamefont{T.}~\bibnamefont{Gaber}},
  \bibinfo{author}{\bibfnamefont{M.}~\bibnamefont{Jetter}},
  \bibinfo{author}{\bibfnamefont{M.}~\bibnamefont{Scheffler}},
  \bibinfo{author}{\bibfnamefont{P.}~\bibnamefont{Michler}},
  \bibinfo{author}{\bibfnamefont{M.}~\bibnamefont{Dressel}},
  \bibinfo{author}{\bibfnamefont{D.}~\bibnamefont{Koelle}},
  \bibnamefont{and} \bibinfo{author}{\bibfnamefont{R.} \bibnamefont{Kleiner}},
  \bibinfo{journal}{Appl. Phys. Lett.} \textbf{\bibinfo{volume}{100}},
  \bibinfo{pages}{012601} (\bibinfo{year}{2012}).
  
\bibitem[{\citenamefont{Gonzalez et~al.}(2007)\citenamefont{Gonzalez, Nunez, Anguita, and Vicent}}]{Gonzalez07}
\bibinfo{author}{\bibfnamefont{E.~M.}~\bibnamefont{Gonzalez}},
  \bibinfo{author}{\bibfnamefont{N.~O.}~\bibnamefont{Nunez}},
  \bibinfo{author}{\bibfnamefont{J.~V.}~\bibnamefont{Anguita}}, \bibnamefont{and}
  \bibinfo{author}{\bibfnamefont{J.~L.} \bibnamefont{Vicent}},
  \bibinfo{journal}{Appl. Phys. Lett.} \textbf{\bibinfo{volume}{91}},
  \bibinfo{pages}{062505} (\bibinfo{year}{2007}).
  
\bibitem[{\citenamefont{Silhanek et~al.}(2008)\citenamefont{Silhanek, Van de Vondel, Moshchalkov, Leo, Metlushko, Ilic, Misko, and Peeters}}]{Silhanek08}
\bibinfo{author}{\bibfnamefont{A.~V.}~\bibnamefont{Silhanek}},
  \bibinfo{author}{\bibfnamefont{J.}~\bibnamefont{Van de Vondel}},
  \bibinfo{author}{\bibfnamefont{V.~V.}~\bibnamefont{Moshchalkov}},
  \bibinfo{author}{\bibfnamefont{A.}~\bibnamefont{Leo}},
  \bibinfo{author}{\bibfnamefont{V.}~\bibnamefont{Metlushko}},
  \bibinfo{author}{\bibfnamefont{B.}~\bibnamefont{Ilic}},
  \bibinfo{author}{\bibfnamefont{V.~R.}~\bibnamefont{Misko}}, \bibnamefont{and}
  \bibinfo{author}{\bibfnamefont{F.~M.} \bibnamefont{Peeters}},
  \bibinfo{journal}{Appl. Phys. Lett.} \textbf{\bibinfo{volume}{92}},
  \bibinfo{pages}{176101} (\bibinfo{year}{2008}).  
  
\bibitem[{\citenamefont{Gonzalez et~al.}(2008)\citenamefont{Gonzalez, Nunez, Anguita, and Vicent}}]{Gonzalez08}
\bibinfo{author}{\bibfnamefont{E.~M.}~\bibnamefont{Gonzalez}},
  \bibinfo{author}{\bibfnamefont{N.~O.}~\bibnamefont{Nunez}},
  \bibinfo{author}{\bibfnamefont{J.~V.}~\bibnamefont{Anguita}}, \bibnamefont{and}
  \bibinfo{author}{\bibfnamefont{J.~L.} \bibnamefont{Vicent}},
  \bibinfo{journal}{Appl. Phys. Lett.} \textbf{\bibinfo{volume}{92}},
  \bibinfo{pages}{176102} (\bibinfo{year}{2008}).

\bibitem[{\citenamefont{Silhanek et~al.}(2005)\citenamefont{Silhanek, Van Look, Jonckheere, Zhu, Raedts, and Moshchalkov}}]{Silhanek05}
\bibinfo{author}{\bibfnamefont{A.~V.}~\bibnamefont{Silhanek}},
  \bibinfo{author}{\bibfnamefont{L.}~\bibnamefont{Van Look}},
  \bibinfo{author}{\bibfnamefont{R.}~\bibnamefont{Jonckheere}},
  \bibinfo{author}{\bibfnamefont{B.~Y.}~\bibnamefont{Zhu}},
  \bibinfo{author}{\bibfnamefont{S.}~\bibnamefont{Raedts}}, \bibnamefont{and}
  \bibinfo{author}{\bibfnamefont{V.~V.}~\bibnamefont{Moshchalkov}},
  \bibinfo{journal}{Phys. Rev. B} \textbf{\bibinfo{volume}{72}},
  \bibinfo{pages}{014507} (\bibinfo{year}{2005}).
  
\bibitem[{\citenamefont{Silhanek et~al.}(2006)\citenamefont{Berdiyorov, Milo\v{s}evi\'{c}, and Peeters}}]{Berdiyorov06}
\bibinfo{author}{\bibfnamefont{G.~R.}~\bibnamefont{Berdiyorov}},
  \bibinfo{author}{\bibfnamefont{M.~V.}~\bibnamefont{Milo\v{s}evi\'{c}}}, \bibnamefont{and}
  \bibinfo{author}{\bibfnamefont{F.~M.}~\bibnamefont{Peeters}},
  \bibinfo{journal}{Phys. Rev. B} \textbf{\bibinfo{volume}{74}},
  \bibinfo{pages}{174512} (\bibinfo{year}{2006}).

\bibitem[{\citenamefont{Rablen et~al.}(2011)\citenamefont{Rablen, Kemmler, Quaglio, Kleiner, Koelle, and Grigorieva}}]{Rablen11}
\bibinfo{author}{\bibfnamefont{S.}~\bibnamefont{Rablen}},
  \bibinfo{author}{\bibfnamefont{M.}~\bibnamefont{Kemmler}},
  \bibinfo{author}{\bibfnamefont{T.}~\bibnamefont{Quaglio}},
  \bibinfo{author}{\bibfnamefont{R.}~\bibnamefont{Kleiner}},
  \bibinfo{author}{\bibfnamefont{D.}~\bibnamefont{Koelle}}, \bibnamefont{and}
  \bibinfo{author}{\bibfnamefont{I.~V.}~\bibnamefont{Grigorieva}},
  \bibinfo{journal}{Phys. Rev. B} \textbf{\bibinfo{volume}{84}},
  \bibinfo{pages}{184520} (\bibinfo{year}{2011}).

\bibitem[{\citenamefont{Cao et~al.}(2011)\citenamefont{Cao, Horng, Wu, Lin, Wu, and Kol\'{a}\v(c)ek}}]{Cao11}
\bibinfo{author}{\bibfnamefont{R.}~\bibnamefont{Cao}},
  \bibinfo{author}{\bibfnamefont{Lance}~\bibnamefont{Horng}},
  \bibinfo{author}{\bibfnamefont{T.~C.}~\bibnamefont{Wu}},
  \bibinfo{author}{\bibfnamefont{J.~C.}~\bibnamefont{Lin}},
  \bibinfo{author}{\bibfnamefont{J.~C.}~\bibnamefont{Wu}},
  \bibinfo{author}{\bibfnamefont{T.~J.}~\bibnamefont{Yang}}, \bibnamefont{and}
  \bibinfo{author}{\bibfnamefont{J.}~\bibnamefont{Kol\'{a}\v(c)ek}},
  \bibinfo{journal}{J. Appl. Phys.} \textbf{\bibinfo{volume}{109}},
  \bibinfo{pages}{083920} (\bibinfo{year}{2011}).

\bibitem[{\citenamefont{Latimer et~al.}(2012)\citenamefont{Latimer, Berdiyorov, Xiao, Kwok, and Peeters}}]{Latimer12}
\bibinfo{author}{\bibfnamefont{M.~L.}~\bibnamefont{Latimer}},
  \bibinfo{author}{\bibfnamefont{G.~R.}~\bibnamefont{Berdiyorov}},
  \bibinfo{author}{\bibfnamefont{Z.~L.}~\bibnamefont{Xiao}},
  \bibinfo{author}{\bibfnamefont{W.~K.}~\bibnamefont{Kwok}}, \bibnamefont{and}
  \bibinfo{author}{\bibfnamefont{F.~M.}~\bibnamefont{Peeters}},
  \bibinfo{journal}{Phys. Rev. B} \textbf{\bibinfo{volume}{85}},
  \bibinfo{pages}{012505} (\bibinfo{year}{2012}).

\bibitem[{\citenamefont{Little and Parks}(1962)\citenamefont{Little and Parks}}]{Little62}
\bibinfo{author}{\bibfnamefont{W.~A.}~\bibnamefont{Little}} \bibnamefont{and}
  \bibinfo{author}{\bibfnamefont{R.~D.} \bibnamefont{Parks}},
  \bibinfo{journal}{Phys. Rev. Lett.} \textbf{\bibinfo{volume}{9}},
  \bibinfo{pages}{9} (\bibinfo{year}{1962}).
  
\bibitem[{\citenamefont{Pannetier et~al.}(1984)\citenamefont{Pannetier, Chaussy, Rammal, and Villegier}}]{Pannetier84}
\bibinfo{author}{\bibfnamefont{B.}~\bibnamefont{Pannetier}}
  \bibinfo{author}{\bibfnamefont{J.}~\bibnamefont{Chaussy}}
  \bibinfo{author}{\bibfnamefont{R.}~\bibnamefont{Rammal}} \bibnamefont{and}
  \bibinfo{author}{\bibfnamefont{J.~C.} \bibnamefont{Villegier}},
  \bibinfo{journal}{Phys. Rev. Lett.} \textbf{\bibinfo{volume}{53}},
  \bibinfo{pages}{1845} (\bibinfo{year}{1984}).
  
\bibitem[{\citenamefont{Behrooz et~al.}(1986)\citenamefont{Behrooz, Burns, Deckman, Levine, Whitehead and Chaikin}}]{Behrooz86}
\bibinfo{author}{\bibfnamefont{A.}~\bibnamefont{Behrooz}}
  \bibinfo{author}{\bibfnamefont{M.~J.}~\bibnamefont{Burns}}
  \bibinfo{author}{\bibfnamefont{H.}~\bibnamefont{Deckman}}
  \bibinfo{author}{\bibfnamefont{D.}~\bibnamefont{Levine}}
  \bibinfo{author}{\bibfnamefont{B.}~\bibnamefont{Whitehead}} \bibnamefont{and}
  \bibinfo{author}{\bibfnamefont{P.~M.}~\bibnamefont{Chaikin}},
  \bibinfo{journal}{Phys. Rev. Lett.} \textbf{\bibinfo{volume}{57}},
  \bibinfo{pages}{368-371} (\bibinfo{year}{1986}).

\bibitem[{\citenamefont{Nori et~al.}(1987)\citenamefont{Nori, Niu, Fradkin and Chang}}]{Nori87}
\bibinfo{author}{\bibfnamefont{F.}~\bibnamefont{Nori}},
  \bibinfo{author}{\bibfnamefont{Q.}~\bibnamefont{Niu}},
  \bibinfo{author}{\bibfnamefont{E.}~\bibnamefont{Fradkin}}, \bibnamefont{and}
  \bibinfo{author}{\bibfnamefont{S.-J.}~\bibnamefont{Chang}},
  \bibinfo{journal}{Phys. Rev. B} \textbf{\bibinfo{volume}{36}},
  \bibinfo{pages}{8338} (\bibinfo{year}{1987}).
  
\bibitem[{\citenamefont{Patel et~al.}(2007)\citenamefont{Patel, Xiao, Hua, Xu, Rosenmann, Novosad,
 Pearson, Welp, Kwok and Crabtree}}]{Patel07}
\bibinfo{author}{\bibfnamefont{U.} \bibnamefont{Patel}},
  \bibinfo{author}{\bibfnamefont{Z.~L.}~\bibnamefont{Xiao}},
  \bibinfo{author}{\bibfnamefont{J.} \bibnamefont{Hua}},
  \bibinfo{author}{\bibfnamefont{T.}~\bibnamefont{Xu}},
  \bibinfo{author}{\bibfnamefont{D.}~\bibnamefont{Rosenmann}},
  \bibinfo{author}{\bibfnamefont{V.}~\bibnamefont{Novosad}},
  \bibinfo{author}{\bibfnamefont{J.}~\bibnamefont{Pearson}},
  \bibinfo{author}{\bibfnamefont{U.}~\bibnamefont{Welp}},
  \bibinfo{author}{\bibfnamefont{W.~K.}~\bibnamefont{Kwok}},
  \bibnamefont{and}
  \bibinfo{author}{\bibfnamefont{G.~W.}~\bibnamefont{Crabtree}},
  \bibinfo{journal}{Phys. Rev. B} \textbf{\bibinfo{volume}{76}},
  \bibinfo{pages}{020508(R)} (\bibinfo{year}{2007}).
  
\bibitem[{\citenamefont{Tinkham}(1996)\citenamefont{Tinkham}}]{Tinkham}
\bibinfo{author}{\bibfnamefont{M.}~\bibnamefont{Tinkham}},
  \bibinfo{journal}{\textit{Introduction to Superconductivity}}
  (\bibinfo{year}{McGraw-Hill, NewYork, 1996}).

\end{thebibliography}
\end{document}